\documentclass{jfm}

\usepackage{graphicx,footmisc}
\usepackage{subfig,epstopdf}
\usepackage{natbib}
\usepackage{lineno,amsmath}

\ifCUPmtlplainloaded \else
  \checkfont{eurm10}
  \iffontfound
    \IfFileExists{upmath.sty}
      {\typeout{^^JFound AMS Euler Roman fonts on the system,
                   using the 'upmath' package.^^J}%
       \usepackage{upmath}}
      {\typeout{^^JFound AMS Euler Roman fonts on the system, but you
                   dont seem to have the}%
       \typeout{'upmath' package installed. JFM.cls can take advantage
                 of these fonts,^^Jif you use 'upmath' package.^^J}%
      }
  \else
  \fi
\fi

\ifCUPmtlplainloaded \else
  \checkfont{msam10}
  \iffontfound
    \IfFileExists{amssymb.sty}
      {\typeout{^^JFound AMS Symbol fonts on the system, using the
                'amssymb' package.^^J}%
       \usepackage{amssymb}%
         
         \let\geq=\geqslant
      }{}
  \fi
\fi

\ifCUPmtlplainloaded \else
  \IfFileExists{amsbsy.sty}
    {\typeout{^^JFound the 'amsbsy' package on the system, using it.^^J}%
     \usepackage{amsbsy}}
    {\providecommand\boldsymbol[1]{\mbox{\boldmath $##1$}}}
\fi

\newcommand\Rey{\mbox{\textit{Re}}}  
\newcommand\Ros{\mbox{\textit{Ro}}}
\newcommand\Pen{\mbox{\textit{Pe}}}  

\newsavebox{\astrutbox}
\sbox{\astrutbox}{\rule[-5pt]{0pt}{20pt}}

\newcommand*{\be}{\begin{equation}}
\newcommand*{\ee}{\end{equation}}

\def\begineq{\begin{equation}}
\def\endeq{\end{equation}}

\def\begineqn{\begin{equation*}}
\def\endeqn{\end{equation*}}

\def\beginar{\begin{eqnarray}}
\def\endar{\end{eqnarray}}
\def\beginarn{\begin{eqnarray*}}
\def\endarn{\end{eqnarray*}}

\def\lb{\left ( }
\def\rb{\right ) }
\def\lsq{\left [ }
\def\rsq{\right ] }

\def\ep{\epsilon}

\def\Rat{\widetilde{Ra}}
\def\Rmt{\widetilde{Rm}}
\def\Pmt{\widetilde{Pm}}

\def\ub{\mathbf{u}}
\def\Bb{\mathbf{B}}
\def\ubp{\mathbf{u}^{\prime}}
\def\bbp{\mathbf{b}^{\prime}}
\def\thp{\theta^{\prime}}

\def\mub{\overline{\bf u}}
\def\mBb{\overline{\bf B}}

\def\mP{\overline{p}}
\def\mPsi{\overline{\Psi}}
\def\mth{\overline{\theta}}
\def\pth{\theta^{\prime}}
\def\pP{p^{\prime}}

\def\mBx{\overline{B}^{(x)}}
\def\mBy{\overline{B}^{(y)}}
\def\mBz{\overline{B}^{(z)}}
\def\Bx{B^{(x)}}
\def\By{B^{(y)}}
\def\Bz{B^{(z)}}
\def\mfu{\overline{u}}
\def\mfv{\overline{v}}
\def\mfw{\overline{w}}
\def\pfu{u^{\prime}}
\def\pfv{v^{\prime}}
\def\pfw{w^{\prime}}
\def\pbx{b^{\prime(x)}}

\def\pby{b^{\prime(y)}}
\def\pbz{b^{\prime(z)}}

\def\dt{{\partial_{T}}}
\def\dtau{{\partial_{\tau}}}
\def\dta{{\partial_\tau}}
\def\dx{{\partial_X}}
\def\dy{{\partial_Y}}

\def\dsx{{\partial_x}}
\def\dsy{{\partial_y}}

\def\dst{{\partial_t}}

\def\dsz{{\partial_z}}

\def\dz{{\partial_Z}}
\def\dzt{{\partial^2_Z}}

\def\hz{{\bf\widehat z}}

\def\hy{{\bf\widehat y}}
\def\hx{{\bf\widehat x}}

\def\litx{{\bf x}}
\def\bigx{{\bf X}}
\def\oAX{\frac{1}{A_X}}
\def\oAZ{\frac{1}{A_Z}}

\def\oAT{\frac{1}{A_T}}
\def\oAt{\frac{1}{A_\tau}}

\def\nabx{{\nabla_{X}}}
\def\nabl{\overline{\nabla}}

\def\cnabx{\cdot{\nabla_{X}}}

\def\div{{\nabla \cdot}}

\def\divl{{\overline{\nabla} \cdot}}
\def\divx{{\nabla_X \cdot}}
\def\lp{{\nabla_\perp^2}}
\def\lap{{\nabla^2}}

\def\wG{\widetilde{\Gamma}}

\newbox\grsign \setbox\grsign=\hbox{$>$} \newdimen\grdimen
\grdimen=\ht\grsign
\newbox\simlessbox \newbox\simgreatbox
\setbox\simgreatbox=\hbox{\raise.5ex\hbox{$>$}\llap
     {\lower.5ex\hbox{$\sim$}}}\ht1=\grdimen\dp1=0pt
\setbox\simlessbox=\hbox{\raise.5ex\hbox{$<$}\llap
     {\lower.5ex\hbox{$\sim$}}}\ht2=\grdimen\dp2=0pt

\title[Multiscale dynamo model]{A multiscale dynamo model driven by quasi-geostrophic convection}

\author[M. A. Calkins, K. Julien, S. M. Tobias and J. M. Aurnou]%
{Michael A. Calkins$^1$
 \thanks{Email address for correspondence: michael.calkins@colorado.edu},\ns
Keith Julien$^1$,
Steven M. Tobias$^2$,
and Jonathan M. Aurnou$^3$}

\affiliation{
$^1$Department of Applied Mathematics, University of Colorado, Boulder, CO  80309, USA \\
$^2$Department of Applied Mathematics, University of Leeds, Leeds, UK LS2 9JT \\
$^3$Department of Earth, Planetary and Space Sciences, University of California, Los Angeles, CA  90095, USA}

\pubyear{2010}
\volume{650}
\pagerange{119--126}
\date{?; revised ?; accepted ?. - To be entered by editorial office}

\begin{document}

\maketitle

\begin{abstract}
A convection-driven multiscale dynamo model is developed in the limit of low Rossby number for the plane layer geometry in which the gravity and rotation vectors are aligned.  The small-scale fluctuating dynamics are described by a magnetically-modified quasi-geostrophic equation set, and the large-scale mean dynamics are governed by a diagnostic thermal wind balance.  The model utilizes three timescales that respectively characterize the convective timescale, the large-scale magnetic evolution timescale, and the large-scale thermal evolution timescale.  Distinct equations are derived for the cases of order one and low magnetic Prandtl number. It is shown that the low magnetic Prandtl number model is characterized by a magnetic to kinetic energy ratio that is asymptotically large, with ohmic dissipation dominating viscous dissipation on the large-scales.  For the order one magnetic Prandtl number model the magnetic and kinetic energies are equipartitioned and both ohmic and viscous dissipation are weak on the large-scales; large-scale ohmic dissipation occurs in thin magnetic boundary layers adjacent to the solid boundaries.  For both magnetic Prandtl number cases the Elsasser number is small since the Lorentz force does not enter the leading order force balance.  The new models can be considered fully nonlinear, generalized versions of the dynamo model originally developed by Childress and Soward [Phys. Rev. Lett., \textbf{29}, p.837, 1972].  These models may be useful for understanding the dynamics of convection-driven dynamos in regimes that are only just becoming accessible to direct numerical simulations. 
\end{abstract}


\section{Introduction}

It is now generally accepted that many of the observed planetary and stellar magnetic fields are the result of convectively driven dynamos \citep{mM05,sS10}.  Direct numerical simulations (DNS) of the complete set of governing equations are now routine practice, but remain limited to parameter values that are quite distant from natural systems owing to the massive requirements for numerically resolving disparate spatiotemporal scales \citep[e.g.][]{cJ11b}.  The stiff character of the governing equations, while an impediment to DNS, provides a possible path forward for simplifying, or reducing, the governing equations with the use of multiscale asymptotics \citep[e.g.][]{kJ07,rK10b}.  Balanced flows, in which two or more forces in the momentum equations are in balance, are particularly suitable for asymptotic analysis given the subdominance of inertial accelerations.  Indeed, reduced models based on the geostrophic balance, in which the Coriolis and pressure gradient forces balance, have formed the backbone for theoretical and numerical investigations on the large-scale dynamics of the Earth's atmosphere and oceans for over 60 years \citep{jC48,jP87}.  


It is an unfortunate fact, however, that direct measurements of the forces present within the electrically conducting regions of natural systems are not possible.  That most large-scale planetary and stellar magnetic fields are aligned with their respective rotation axes \citep[e.g.][]{gS11} suggests that the Coriolis force plays a key role in the magnetic field generation process, at least with respect to the large-scale dynamics.  For the case of the Earth's liquid outer core, viscous stresses are likely to be small for large-scale motions there \citep{mP13}, and observations tracking the movement of the geomagnetic field show that typical convective timescales are significantly longer than the rotation period \citep{cF11}.  These studies suggest that the large-scale dynamics within the core may be geostrophically balanced \citep{dJ88,aJ93,nG10,nS11,nG12}.  Alternatively, it has been hypothesized that the Lorentz force can balance with the Coriolis and pressure gradient forces to yield magnetostrophically balanced motions at certain lengthscales within the core \citep{pR88a,hM08}.  It can be argued that only geostrophically balanced dynamos have been observed with DNS \citep[e.g.][]{kS12,eK13b} (However, see the plane layer investigation of \cite{jR02}).

In a seminal investigation, \cite{sC72} demonstrated that laminar convection in a rotating plane layer geometry is capable of supporting dynamo action.  They outlined three unique distinguished limits, or balances, in the governing equations that can occur based on the relative strength of the magnetic field, which they referred to as the weak field, intermediate field, and strong field limits.  \cite{aS74} (denoted S74 hereafter) and \cite{yF82} (denoted FC82 hereafter) subsequently investigated the weak and intermediate field limits in greater detail; for both investigations the flows were weakly nonlinear.  S74 showed that stable periodic dynamos exist in the weak field limit, whereas \cite{yF82} found that dynamos of intermediate field strength may be dynamically unstable and thus lead to strong field states as the flow saturates.  For simplicity, we refer to both the S74 and FC82 cases as the Childress-Soward dynamo model (denoted by CSDM), since they were first discussed by \cite{sC72} and the only difference between the two models is the strength of the magnetic field.  The basis for the CSDM is that the Ekman number $E_H = \nu / 2\Omega H^2$ is taken as a small parameter, where $\nu$ is the kinematic viscosity, $\Omega$ is the rotation rate and $H$ is the depth of the fluid layer. In this case convection is spatially anisotropic with large aspect ratio $H/\ell \gg 1$, where $\ell$ characterizes the small horizontal length scale of the convection \citep{sC61}.  By expanding the flow variables in powers of $E^{1/6}$, a dynamical equation for the large-scale horizontal magnetic field is derived that is driven by the mean electromotive force (emf) generated by the small-scale fluctuating velocity and magnetic fields.  Because the CSDM utilizes a single time scale, it applies to order one thermal and magnetic Prandtl numbers, $Pr=\nu/\kappa$ and $Pm = \nu / \eta$, respectively, where $\kappa$ is the thermal diffusivity and $\eta$ is the magnetic diffusivity of the fluid. 

The CSDM continues to provide an invaluable tool for understanding the differences between large-scale and small-scale dynamo action, as well as aiding the interpretation of results obtained from DNS studies.  S74 showed that a dynamo driven by small-scale rapidly rotating convection is necessarily large-scale.  Many of the key features predicted by S74, such as the vertical structure and oscillatory temporal evolution of the large-scale magnetic field, have been confirmed with low Ekman number DNS \citep[e.g.][]{sS04}.  Recent work has employed the CSDM for investigating kinematic dynamo action driven by rotating convection \citep{bF13}.  \cite{kM13} have extended the CSDM to include the effects of fluid compressibility via the anelastic approximation, showing that density stratification tends to delay the onset of dynamo action and reduces the strength of the resulting magnetic field.  

In the present work we develop a new multiscale dynamo model that possesses many similarities with the CSDM, but also significant differences.  We refer to the new model as the quasi-geostrophic dynamo model, or QGDM.  Like the CSDM, the small-scale, convective dynamics are geostrophically balanced to leading order, but remain time-dependent and fully nonlinear in the QGDM; our small-scale dynamical equations are a magnetically modified version of the quasi-geostrophic convection equations developed by \cite{kJ98a} (see also \cite{kJ06} for a generalized development).  The strength of the magnetic field in the QGDM is asymptotically larger than that considered by FC82, but is not a strong field in the sense that the Lorentz force does not enter the leading order force balance.  In addition to the small horizontal convective lengthscale $\ell$, the QGDM includes large-scale horizontal modulations $L_X$ that allow for a diagnostic thermal wind balance in the large-scale momentum equations.  Coupled with the large-scale heat equation, the large-scale model (in the absence of a magnetic field) is the equivalent of the planetary geostrophic equations commonly employed in oceanography \citep{aR59,pW59,iG11} and atmospheric science \citep{nP63,sD09}. (The so-called $\beta$-effect has been neglected in the present work though it can easily be incorporated into the current model, e.g.~see \cite{iG11}).  Because of the additional scale $L_X$, the QGDM also possesses a non-zero large-scale vertical magnetic field that is not present in the CSDM.  Moreover, the new model utilizes three disparate timescales characterizing the small-scale convective dynamics, the large-scale magnetic evolution timescale, and the large-scale thermal evolution timescale.  Both low and order one magnetic Prandtl number cases are considered.  The QGDM can be simulated numerically to gain insight into large-scale magnetic field generation in planets and stars, where we can utilize the success of previous non-magnetic work \cite[e.g.][]{mS06,kJ12b,kJ12,aR14,sS14}, and the increases in computational power that have occurred since the work of CS72.  

In section \ref{S:model} we present the asymptotic development of the model.  In section \ref{S:discuss} we discuss some of the important features of the QGDM, its relationship with the CSDM, and possible applications to natural dynamos.  Concluding remarks are given in section \ref{S:conclude}.



\section{Model Development}
\label{S:model}

\subsection{Governing Equations}

We consider a rotating plane layer geometry with rotation vector $\mathbf{\Omega} = \Omega \, \hz$ and constant gravity vector $\boldsymbol{g} = - g \hz$.  The horizontal boundaries are located a vertical distance $H$ apart with the layer heated from the bottom boundary and cooled at the top boundary.  For simplicity, the fluid is assumed to be Newtonian and Boussinesq with density $\rho$ and thermal expansion coefficient $\alpha$.  The governing equations are then non-dimensionalized utilizing the scales $\mathcal{U}$, $\ell$, $\ell/\mathcal{U}$, $\Delta T$, $\mathcal{P}$ and $\mathcal{B}$ for the velocity, length, time, temperature, pressure, and magnetic field, respectively.  In the rotating reference frame with coordinates $(x,y,z)$ the dimensionless governing equations are then
\begin{gather}
\dst \ub + \ub \cdot \nabla \ub + \frac{1}{\Ros} \, \hz \times \ub = - Eu \, \nabla p + M \, \Bb \cdot \nabla \Bb + \Gamma \, \theta \, \hz + \frac{1}{\Rey} \nabla^2 \ub, \label{E:mom1} \\
\dst \theta + \div \lb \ub \theta \rb =  \frac{1}{\Pen} \nabla^2 \theta, \\
\dst \Bb  =  \nabla \times ( \ub \times \Bb) +  \frac{1}{Rm} \nabla^2 \Bb , \\
\div \ub =  0 , \quad \div \Bb  =  0 \label{E:Sol1} ,
\end{gather}
where the velocity, pressure, temperature, and magnetic field are denoted by $\ub = \lb u,v,w \rb$, $p$, $\theta$, and $\Bb = \lb \Bx, \By, 
\Bz \rb$, respectively.  Both the hydrostatic centrifugal force and magnetic pressure have been absorbed in the pressure gradient term $\nabla p$.  The dimensionless parameters are defined by 
\be
Ro = \frac{\mathcal{U}}{2 \Omega \ell}, \quad
{\Gamma}=\frac{g \alpha\Delta T \ell}{\mathcal{U}^2},\quad
Re=\frac{\mathcal{U} \ell}{\nu},\quad  
Pe=\frac{\mathcal{U} \ell}{\kappa},\quad
Eu=\frac{\mathcal{P}}{\rho \, \mathcal{U}^2} , \label{E:params1}
\ee
\be
M = \frac{\mathcal{B}^2}{\rho \mu \, \mathcal{U}^2}, \quad
Rm = \frac{\mathcal{U} \ell}{\eta} . \label{E:params2}
\ee
Here, $Ro$ is the Rossby number, $\Gamma$ is the buoyancy number, $Re$ is the Reynolds number, $Pe$ is the P\'eclet number, $Eu$ is the Euler number, $M$ is the ratio of magnetic energy to kinetic energy, and $Rm$ is the magnetic Reynolds number.  The permeability of free space is denoted by $\mu$.  We recall that the thermal and magnetic Prandtl numbers characterizing the fluid properties are related to the above parameters via the relationships $Pe = Re Pr$ and $Rm = Re Pm$.  Importantly, in the present work we denote dimensionless parameters that are based on the large-scale depth of the fluid layer with a subscript $H$, whereas parameters without this distinction utilize the small convective scale $\ell$.

In what follows we assume that the horizontal bounding surfaces are impenetrable, stress-free and perfect thermal and electrical conductors such that
\begineq
\dsz u = \dsz v = w = \dsz \Bx = \dsz \By = \Bz =0, \quad \textnormal{at} \quad z=0,H/ \ell ,
\endeq
\begineq
\theta = 1 \quad \textnormal{at} \quad z = 0, \quad \textnormal{and} \quad \theta = 0 \quad \textnormal{at} \quad z = H/\ell .
\endeq

\begin{figure}
\begin{center}
  \includegraphics[width=13cm]{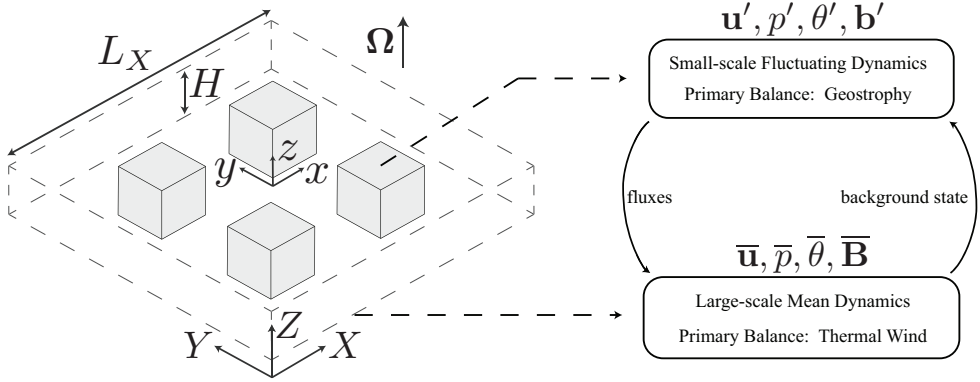}
\caption{Schematic summarizing the general features of the multiscale dynamo model. The large-scale coordinate system is $\mathbf{X} = (X,Y,Z)$ and the small-scale coordinate system is $\mathbf{x} = (x,y,z)$; the shaded boxes depict small-scale subdomains embedded within the single large-scale domain.  Large-scale mean dependent variables are given by $(\mub,\mP,\mth,\mBb)$ and the small-scale fluctuating dependent variables are $(\mathbf{u}', p', \theta', \mathbf{b}')$.  Distinct equation sets are developed for the large-scale and small-scale domains, with the mean dynamics controlling the background state for the fluctuating dynamics and the fluctuating dynamics feeding back onto the mean state via small-scale fluxes.}
\label{F:dynamo}
\end{center}
\end{figure}

The development given here parallels that of \cite{kJ07} and \cite{iG11} for other multiscale models.   We find that three disparate timescales are necessary to completely describe the temporal evolution of the dynamo; these are (1) the timescale of the small-scale convective fluctuations (denoted by $t$), (2) the large-scale magnetic evolution timescale ($\tau$), and (3) the large-scale thermal evolution timescale ($T$).  The small-scale, rapidly varying spatial and temporal coordinates are denoted as ($\litx, t$), and the slowly varying large-scale coordinates are given by ($\bigx, \tau, T$).  The fields are decomposed into mean and fluctuating components according to 
\beginar
\ub (\litx,\bigx,t,\tau,T) & = & \mub(\bigx,\tau,T) + \ubp(\litx,\bigx,t,\tau,T), \\
p(\litx,\bigx,t,\tau,T) & = & \mP (\bigx,\tau,T) + \pP (\litx,\bigx,t,\tau,T), \\
\theta (\litx,\bigx,t,\tau,T) & = & \mth (\bigx,\tau,T) + \pth (\litx,\bigx,t,\tau,T), \\
\Bb (\litx,\bigx,t,\tau,T) & = & \mBb (\bigx,\tau,T) + \bbp (\litx,\bigx,t,\tau,T),
\endar
and the fluctuating magnetic field vector is $\bbp = (\pbx,\pby,\pbz )$.  In Figure \ref{F:dynamo} a schematic illustrating the general features of the model is shown.
The mean quantities are defined as averages taken over the fast scales according to
\begineq
\overline{f}(\bigx,\tau,T)  = \lim_{t', \mathcal{V} \rightarrow \infty} \, \frac{1}{t' \mathcal{V}} \int_{t',\mathcal{V}} f(\litx,\bigx,t,\tau,T) \, d \mathbf{x} \, dt , \quad \overline{f'} \equiv 0,
\endeq
with the small-scale fluid volume denoted by $\mathcal{V}$.
With multiple scales the differentials become
\beginar
\dst & \rightarrow & \dst + \oAt \dta + \oAT \dt , \\
\nabla & \rightarrow & \nabla + \nabl  ,
\endar
where now $\nabla$ only acts on the small-scale coordinates $\bf x$ and the large-scale, mean gradient operator is defined by 
\be
\nabl = \lb \oAX \dx, \oAX \dy, \oAZ \dz \rb .
\ee
The various aspect ratios are defined according to 
\be
A_X = A_Y = \frac{L_X}{\ell}, \quad A_Z = \frac{H}{\ell}, \quad A_\tau = \frac{t}{\tau}, \quad A_T = \frac{t}{T} . \label{E:aspect}
\ee
For simplicity we assume isotropic horizontal large-scale modulations (i.e.~$A_X = A_Y$) in the present work, though the approach remains completely general.

The governing equations become
\begineq
\begin{split}
  \lb \dst + \oAt \dta + \oAT \dt \rb \ub  + \div \lb \ub \ub^{tr} \rb + \divl \lb \ub \ub^{tr} \rb + \frac{1}{\Ros} \, \hz \times \ub   = \\  - Eu \lb \nabla + \nabl\rb p  + M \lb  \Bb \cdot \nabla  + \Bb \cdot \nabl  \rb \Bb + \Gamma \, \theta \hz +  \frac{1}{\Rey} \lb  \nabla^2 + \nabl^2 \rb \ub  , \label{E:mom2}
\end{split}
\endeq
\begineq
\begin{split}
\lb \dst + \oAt \dta + \oAT \dt \rb \theta + \div \lb \theta \ub  \rb + \divl \lb \theta\ub \rb  =   \frac{1}{\Pen} \lb  \nabla^2 + \nabl^2 \rb  \theta ,
\end{split}
\endeq
\begineq 
\begin{split}
\lb \dst + \oAt \dta + \oAT \dt \rb  \Bb = \nabla \times ( \ub \times \Bb) + \nabl \times ( \ub \times \Bb) +   \frac{1}{Rm}\lb  \nabla^2 + \nabl^2 \rb  \Bb , 
\end{split}
\endeq
\begineq
\lb \nabla +  \nabl \rb \cdot \ub  = 0, \quad \lb \nabla + \nabl  \rb \cdot \Bb  = 0 . \label{E:Sol2}
\endeq
The superscript ``$tr$" appearing on some of the velocity vectors denotes a transpose.

Averaging the equations over fast temporal and spatial scales $(\litx,t)$ results in the mean equations
\begineq
\begin{split}
\lb \oAt \dtau + \oAT \dt \rb \mub + \divl \lb \mub (\mub)^{tr} \rb  + \divl \overline{\lb \ubp (\ubp)^{tr} \rb} +   \frac{1}{\Ros} \, \hz \times \mub  = \\  - Eu \nabl \mP + M \overline{\mathbf{F}}_L + \Gamma \, \overline{\theta} \, \hz + \frac{1}{\Rey} \nabl^2 \mub ,
\label{E:meanmom}
\end{split}
\endeq
\beginar
\lb \oAt \dtau + \oAT \dt \rb \mth + \divl \lb \mth \, \mub \rb +  \divl \overline{\lb \pth \ubp  \rb} & = & \frac{1}{\Pen} \nabl^2\mth \label{E:meantemp} , \\
\lb \oAt \dtau + \oAT \dt \rb \mBb = \nabl \times ( \mub \times \mBb )  +  \nabl \times \overline{ (\ubp \times \bbp)} & + & \frac{1}{Rm} \nabl^2 \mBb \label{E:minduc0},
\endar
\be
\nabl \cdot \mub = 0, \quad \nabl \cdot \mBb = 0,
\ee
where the mean Lorentz force is $\overline{\mathbf{F}}_L = \mBb \cdot \nabl \,\mBb + \overline{\bbp \cdot \nabla \bbp}$.  

The fluctuating equations are found by subtracting the mean equations from the full equations to give 
\begineq
\begin{split}
\lb \dst + \oAt \dta + \oAT \dt \rb \ubp  + \mub \cdot \nabla \ubp + \ubp \cdot \nabl \ub - \nabl \cdot \overline{\lb \ubp (\ubp)^{tr} \rb}   + \frac{1}{\Ros} \, \hz \times \ubp = \\  - Eu \, \lb \nabl + \nabla \rb \pP + M \, \mathbf{F}_L^{\prime}+ \Gamma \, \thp \, \hz + \frac{1}{\Rey} \lb \nabla^2 + \nabl^2 \rb \ubp  ,
\label{E:flucmom}
\end{split}
\endeq
\begineq
\begin{split}
\lb \dst + \oAt \dta + \oAT \dt \rb \pth +   \ubp \cdot \nabl \, \mth  + \mub \cdot \nabla \pth +  \nabl \cdot \lb \ubp \pth \rb - \nabl \cdot \overline{ \lb \ubp \pth \rb}    = \\  \frac{1}{\Pen} \lb  \nabla^2 + \nabl^2 \rb \pth ,
\end{split}
\endeq
\begineq
\begin{split}
\lb \dst + \oAt \dta + \oAT \dt \rb \bbp  = \\  \nabl \times ( \mub \times \bbp) +  \nabl \times ( \ubp \times \mBb) +   \nabl \times ( \ubp \times \bbp) - \nabl \times \overline{(\ubp \times \bbp)} + \\ \nabla \times ( \mub \times \bbp)  +   \nabla \times ( \ubp \times \mBb)  +  \nabla \times ( \ubp \times \bbp)  + \frac{1}{Rm} \lb \nabla^2 + \nabl^2 \rb \bbp \label{E:finduc0} ,
\end{split}
\endeq
\begineq
\lb \nabl + \nabla \rb \cdot \ubp = 0, \quad \lb \nabl + \nabla \rb \cdot \bbp = 0 , \label{E:flucsolen}
\endeq
where the fluctuating Lorentz force is $\mathbf{F}_L^{\prime} = \mBb \cdot \nabla \bbp + \bbp \cdot \nabl \, \mBb + \bbp \cdot \nabla \bbp - \overline{\bbp \cdot \nabla \bbp}$.

\subsection{Asymptotics}

We note that up to this point, no approximations have been made; the equations have simply been split up into mean and fluctuating components.  In the present work we are interested in the development of a multiscale dynamo model that preserves geostrophic balance on the small (fluctuating) scales.  Because the fluctuating and mean dynamics are coupled, \textit{a posteriori} we find that the large-scales are required to also be geostrophically balanced in the horizontal directions, and hydrostatic in the vertical direction, resulting in a thermal wind balance.  The relevant asymptotic limit for the present model is therefore the small-Rossby number limit, $Ro \equiv \ep \rightarrow 0$.  

When constructing an asymptotic model with more than one small (or large) parameter, it is necessary to establish the so-called distinguished limits, or the asymptotic scaling relationships between the various parameters.  In the present context this implies that we must relate, in an order of magnitude sense, both the non-dimensional parameters introduced in equations \eqref{E:params1}-\eqref{E:params2}, as well as the aspect ratios defined by equations \eqref{E:aspect}, to the Rossby number $\ep$.  This procedure is well known, and hinges on the identification of dominant balances in the governing equations that will allow for a mathematically and physically meaningful reduced model; by definition the procedure is circular in nature \citep[e.g.~see][]{cB10}.  In this regard, to ensure geostrophically balanced convection on the small scales, we follow section 2 of \cite{mS06} and employ the following distinguished limits
\begineq
A_Z = \ep^{-1},  \quad A_T = \ep^{-2}, \quad Eu = \ep^{-2}, \quad \Gamma = \ep^{-1} \widetilde{\Gamma} ,  \quad Re = O(1), \quad Pe = O(1),
\endeq
where the reduced buoyancy number $ \widetilde{\Gamma}=O(1)$.  We note that with these scalings, the small-scale velocity field $\ubp$ is $O(1)$.  We further assume the large-scale pressure $\mP$, magnetic field $\mBb$ and temperature $\mth$ to also be $O(1)$ in magnitude.  

The dimensional scales employed in the dimensionless parameters given by equations \eqref{E:params1}-\eqref{E:params2} will be those of the largest amplitude.  For convenience the parameters $Eu$, $\Gamma$, and $M$ will therefore be based on `mixed' scales in our asymptotic development in the sense that the pressure, temperature, and magnetic field amplitudes are based on the large-scale quantities ($\mP,\mBb,\mth$) whereas the velocity scale is based on $\ubp$.  In this sense, the dimensional magnitudes of these variables is absorbed in the various non-dimensional parameters. In section \ref{S:apps} we discuss the renormalization of these non-dimensional parameters such that they are based on equivalent (large or small) dimensional scales.

It now remains to determine the distinguished limits of $A_X$, $A_\tau$, $M$, and $Rm$.  To provide some explanation as to how these limits are achieved we examine various terms in the governing equations that we wish to retain in the final reduced model.  The distinguished limit of $A_X$ can be determined by restricting the large-scale horizontal motions to be geostrophically balanced; utilizing equation \eqref{E:meanmom} we have
\be
\frac{1}{\ep} \hz \times \mub \approx - \frac{1}{\ep^2 A_X} \nabx \mP, 
\ee
where $\nabx = (\dx , \dy, 0)$. We then get 
\be
\mub \sim \lb \ep A_X \rb^{-1} \label{E:us1} .
\ee
Turning our attention to the mean temperature equation \eqref{E:meantemp}, we wish to keep horizontal advection of the mean temperature by the mean velocity field at the same order as the temporal evolution of the mean temperature so that
\be
\ep^2 \dt \mth \sim \frac{1}{A_X} \mub \cdot \nabx \mth,
\ee
which leads to 
\be
\mub \sim \ep^2 A_X . \label{E:us2}
\ee
Combining \eqref{E:us1} with \eqref{E:us2} then shows that we must have
\be
A_X = \ep^{-3/2} ,
\ee
and so $\mub = O(\ep^{1/2})$.  We note that this also implies large-scale horizontal gradients are smaller than the corresponding large-scale vertical gradients by a factor $A_Z/A_X=O(\ep^{1/2})$.

The distinguished limits of $A_\tau$, $M$, and $Rm$ are determined by examining the influence of the magnetic field.  To generate dynamo action on the large scales via coupling with the fluctuating dynamics, we necessarily require the presence of the second term on the righthand side of \eqref{E:minduc0}, where $\overline{(\ubp \times \bbp)}$ is referred to as the mean electromotive force, or mean emf \citep[e.g.][]{eP55,mS66a,mS66b,hM78b}.  Utilizing the above scalings for $A_X$ and $\mub$, and retaining only the largest terms, the order of magnitude of each term present in the mean induction equation is then
\be
\frac{\mBb}{A_\tau} \hspace{0.2cm} : \hspace{0.2cm} \ep^{3/2} \mBb \hspace{0.2cm} : \hspace{0.2cm} \ep \bbp \hspace{0.2cm} : \hspace{0.2cm} \frac{\ep^2}{Rm} \mBb ,  \label{E:mord}
\ee
where we recall that the fluctuating velocity field is order one, consistent with \cite{mS06}.  The colons used above are meant to signify that we are making a comparison of the magnitude of each term in the original governing equation and we have retained $\mBb = O(1)$ for clarity in identifying the various terms.  To retain time dependence of the mean magnetic field, we require that the first and third terms be of the same magnitude such that
\be
\mBb \sim \ep A_\tau \bbp . \label{E:Bt}
\ee
By noting that all the large-scale gradients are small relative to the small-scale gradients, the order of magnitude of the four largest terms (the first on the lefthand side, and the fifth, sixth and eighth terms on the righthand side) in the fluctuating induction equation \eqref{E:finduc0} are 
\be
\bbp \hspace{0.2cm} : \hspace{0.2cm} \mBb \hspace{0.2cm} : \hspace{0.2cm} \bbp \hspace{0.2cm} : \hspace{0.2cm} \frac{\bbp}{Rm} . 
\ee
We require that ohmic dissipation is present in the final model, at least with respect to the small scales.  The largest term present above would then be that associated with stretching the mean magnetic field such that
\be
\mBb \sim \frac{\bbp}{Rm} ,  \label{E:Bs1}
\ee
i.e.~$\bbp \sim Rm$. Finally, because we're investigating dynamo action, we require that the Lorentz force enter the small scale momentum dynamics.  From \cite{mS06}, and examination of equation \eqref{E:flucmom}, we know that this requires 
\be
M \mBb \bbp = O(1) .
\ee
Combining this scaling with \eqref{E:Bs1} yields
\be
M = O(Rm^{-1}) . \label{E:Ms}
\ee
We can now determine the size of $Rm$ by returning to the mean induction equation.  For large-scale dynamo action, magnetic diffusion cannot dominate the mean emf so that comparison of the third and fourth terms, respectively, given in \eqref{E:mord} yields
\be
\ep \bbp \gtrsim \frac{\ep^2}{Rm} \mBb .
\ee
Using the above scaling with \eqref{E:Bs1} shows that 
\be
Rm \geq \ep^{1/2},
\ee
with the lower bound being the case $Rm=O(\ep^{1/2})$.  The particular distinguished limit taken for $Rm$ will in turn determine the size of the magnetic Prandtl number since $Rm = Re Pm$.  Taking $Rm=O(\ep^{1/2})$ corresponds to the low $Pm$ limit since $Re = O(1)$.  An alternative scaling would be to take $Rm = O(1)$ and thus $Pm = O(1)$.  The main focus of the present manuscript is for the $Rm = O(\ep^{1/2})$ limit given that it ties directly to the CSDM and is implicitly low $Pm$.  We also present the form of the governing equations for the $Rm = O(1)$ case, but note additional complications that arise in this limit which will require future analysis that is beyond the scope of the present work. 


\subsubsection{The $Rm = O(\ep^{1/2})$ Limit}

Here we take $Rm = \ep^{1/2} \Rmt$, where we define the reduced magnetic Reynolds number $\Rmt = O(1)$.  Returning to relationships \eqref{E:Bt} and \eqref{E:Ms} we can now define the distinguished limits of $A_\tau$ and $M$ as 
\be
A_\tau = \ep^{-3/2}, \quad M = \ep^{-1/2} ,
\ee
showing that three disparate timescales are now present in the QGDM.  

With our distinguished limits defined, we now follow CS72 and expand all variables in powers of $\ep^{1/2}$, e.g.
\be
\ub = \mub_0 +  \ubp_0 + \ep^{1/2} \lb \mub_{1/2} +  \ubp_{1/2} \rb + \ep \lb \mub_{1} +  \ubp_{1} \rb + \cdots.
\ee
Plugging the perturbation expansions into the governing equations and collecting terms of equal magnitude, the leading order solenoidal conditions on the mean velocity and magnetic fields gives
\begineq
\dz \mfw_0 = 0, \quad \dz \mBz_0 = 0 .
\endeq
Owing to our use of impenetrable, perfectly conducting boundary conditions we require $\mfw_0 = \mBz_0 \equiv 0$.  At the next order the solenoidal conditions yield
\begineq
\divx \mub_0 + \dz \mfw_{1/2} = 0, \quad \divx \mBb_0 + \dz \mBz_{1/2} = 0 . \label{E:solen}
\endeq
At order $\mathcal{O}(\ep^{-1})$ the horizontal mean momentum equation gives
\begineq
\hz \times \mub_0 = 0, 
\endeq
showing that $\mub_0 \equiv 0$. Utilizing continuity it follows that $\mfw_{1/2} \equiv 0$. The next three orders of the mean horizontal momentum equation are then geostrophically balanced,
\be
\hz \times \mub_i = - \nabx \mP_{i-1/2}, \quad i = \frac{1}{2},1,\frac{3}{2}.\label{E:mgeo}
\ee
Carrying the expansion out to $O(\ep)$ shows that $\mub_2$ is magnetostrophically balanced,
\be
\hz \times \mub_{2} = -\nabla_X \mP_{3/2} + \mBb_0 \cnabx \mBb_0 + \mBz_{1/2} \dz \mBb_0 .\label{E:mag}
\ee
The vertical mean momentum equation yields hydrostatic balance for the first four orders, 
\be
\dz \mP_i = \wG \mth_i , \quad i = 0,\frac{1}{2},1,\frac{3}{2} . \label{E:mhydro} 
\ee
Taking the curl of equation \eqref{E:mgeo} shows that 
\be
\nabla_X \cdot \mub_{i} = 0, \quad i = 0,\frac{1}{2},1,\frac{3}{2} , \label{E:mcont} 
\ee
and thus $\mfw_{i} \equiv 0$ for $i=0,\ldots 2$.  Combining equation \eqref{E:mgeo} for $i=1/2$ and \eqref{E:mhydro} for $i=0$ we then obtain the well-known thermal wind relations
\begineq
\dz \mfv_{1/2} =  \wG \dx \mth_0, \quad 
\dz \mfu_{1/2} =  - \wG \dy \mth_0 . \label{E:twind}
\endeq
Because the large-scale velocity field is horizontally divergence free we can define the large-scale geostrophic stream function as $\mPsi_{1/2} \equiv \mP_0$ such that
\be
\mub_{1/2} = - \nabl \times \mPsi_{1/2} \hz .
\ee

We mention that the mathematical structure of the thermal wind relations \eqref{E:twind} yields any barotropic (i.e. depth invariant) geostrophic large-scale flow undetermined given that $\mub_{1/2}$ only appears with first-order $Z$-derivatives.  In some instances this barotropic flow plays a significant dynamical role and must be consistently determined via an evolution equation \citep[c.f.][]{sD09}.  However, in the present work the evolution equation for such a mode enters at a much higher, subdominant order, and we do not consider it any further. 

Proceeding to the mean horizontal induction equation, at $O(\ep)$ we have
\begineq
0 = \dz \lsq \hz \times \overline{\lb \ubp_0 \times \bbp_0  \rb} \rsq , 
\endeq
which, in general, requires that $\bbp_0 \equiv 0$.  Alternatively, from the scaling relationship given by \eqref{E:Bs1}, we require that $\bbp = O(\ep^{1/2})$ if $\mBb = O(1)$.  At $O(\ep^{3/2})$ we get 
 \begineq
 \partial_\tau \mBb^{\perp}_0 =   \dz \lsq \hz \times \overline{\lb \ubp_0 \times \bbp_{1/2}  \rb} \rsq + \frac{1}{\Rmt} \dzt \mBb^{\perp}_0 ,
 \endeq
where $\mBb^{\perp}_0 \equiv (\mBx_0,\mBy_0,0)$.  The leading order vertical component of the mean induction equation becomes
 \begineq
  \partial_\tau \mBz_{1/2} = \dx \lsq \hy \cdot \overline{\lb \ubp_0 \times \bbp_{1/2}  \rb} \rsq - \dy \lsq \hx \cdot \overline{\lb \ubp_0 \times \bbp_{1/2}  \rb} \rsq  + \frac{1}{\Rmt} \dzt \mBz_{1/2},
 \endeq
where we see that a non-trivial mean vertical magnetic field requires the presence of a large-scale horizontal modulation in the sense that $\mBz_{1/2}\equiv 0$ if $\dx=\dy\equiv0$ (this fact can also be seen from equation \eqref{E:solen}).  The fluctuating induction equation then gives 
 \begineq
 0 = \mBb^{\perp}_0 \cdot \nabla_\perp \ubp_0 + \frac{1}{\Rmt} \lap \bbp_{1/2}, \label{E:fmag}
 \endeq
 where $\nabla_\perp = \lb \dsx,\dsy,0 \rb$.  

The leading order mean temperature equation gives
\begineq
\dz \overline{\lb \pfw_0 \pth_0 \rb} = 0,
\endeq
showing that $\pth_0 \equiv 0$.  At the next order we have
\begineq
\partial_\tau \mth_0 + \dz \overline{\lb \pfw_0 \pth_{1/2} \rb} = 0,
\endeq
again showing that $\pth_{1/2} \equiv 0$ and thus $\partial_\tau \mth_0 \equiv 0$.  Finally, we have
\begineq
\partial_\tau \mth_{1/2} + \dt \mth_0 + \mub_{1/2} \cdot \nabla_X \mth_0 + \dz \overline{\lb \pfw_0 \pth_1 \rb} = \frac{1}{Pe} \dzt \mth_0 . \label{E:mht}
\endeq
To avoid secular growth of the mean temperature on the timescale $\tau$ we set $\partial_\tau \mth_{1/2} \equiv 0$; the mean heat equation then becomes 
\begineq
\dt \mth_0 + \mub_{1/2} \cdot \nabla_X \mth_0 + \dz \overline{\lb \pfw_0 \pth_1 \rb} = \frac{1}{Pe} \dzt \mth_0 . \label{E:mht2}
\endeq
Alternatively, one can average equation \eqref{E:mht} over the timescale $\tau$ to obtain an equation identical to \eqref{E:mht2}, with the exception that the averages must then be interpreted as occuring over $\lb \mathbf{x},t,\tau \rb$.  At $O(\ep)$ the fluctuating temperature equation is 
\be
\dst \pth_1  + \ubp_0 \cdot \nabla_\perp \pth_1 + \pfw_0 \dz \mth_0 = \frac{1}{Pe} \lap \pth_1 .
\ee

At $O(\ep^{-2})$ and $O(\ep^{-3/2})$ the fluctuating momentum equation yields, respectively
\begineq
\nabla \pP_i = 0, \quad i = 0, \frac{1}{2} ,
\endeq
showing that $\pP_0 = \pP_{1/2} \equiv 0$.  Geostrophy occurs at $O(\ep^{-1})$ and $O(\ep^{-1/2})$
\begineq
\hz \times \ubp_i = - \nabla \pP_{i+1}, \quad i=0,\frac{1}{2} . \label{E:flucgeo}
\endeq
Additionally, mass conservation at $O(1)$ and $O(\ep^{1/2})$ gives
\be
\div \ubp_i = 0, \quad i=0,\frac{1}{2}, \label{E:cont}
\ee
which, along with equations \eqref{E:flucgeo}, yields the Proudman-Taylor theorem acting over the small vertical scale $z$
\be
\dsz \lb  \ubp_i, \pP_{i+1} \rb = 0, \quad i=0,\frac{1}{2} .
\ee

It is natural to wonder why the Proudman-Taylor theorem is satisfied over the small scale $z$ and not the axial domain scale $Z$.  It is well known from the linear theory of rapidly rotating convection that the leading order, geostrophically balanced convection possesses order one variations over the height of the fluid layer (our large-scale coordinate $Z$) \citep{sC61}.  The only way to be consistent with the Proudman-Taylor theorem, whilst allowing convection to occur on the axial domain scale, is for it to be satisfied over the small vertical scale $z$.  We recall that this is the same scale as the horizontal convective scales $(x,y)$.  This fact appears to be first noted by \cite{kS79}; apart from the presence of viscous and body forces in the present work, their vorticity (2.7b) and vertical momentum (2.7a) equations are identical to our small-scale counterparts given by \eqref{E:vort0} and \eqref{E:mom0}.


The prognostic momentum equation then appears at $O(1)$
\begineq
\dst \ubp_0 + \ubp_0 \cdot \nabla_\perp \ubp_0 +  \hz \times \ubp_1 = - \nabla \pP_2 - \dz \pP_1 \hz + \wG \pth_1 \hz + \mBb_0 \cdot \nabla_\perp \bbp_{1/2} + \frac{1}{Re} \lp \ubp_0 , \label{E:momfluc}
\endeq
where we note the key distinction with the CSDM is the simultaneous appearance of both the advection and Lorentz force terms.  In light of equation \eqref{E:flucgeo}, we can define the small-scale geostrophic stream function as $\psi_0' \equiv \pP_1$ such that the fluctuating horizontal velocity field is given by $(\pfu_0,\pfv_0) = (-\dsy \psi_0', \dsx \psi_0')$.  As noted by \cite{mS06}, equation \eqref{E:momfluc} still depends upon the small-scale vertical scale $z$; to avoid secular growth on this scale it is necessary to impose solvability conditions.   These solvability conditions amount to operating on equation \eqref{E:momfluc} with $\hz \cdot \nabla \times$ and $\hz \cdot$, respectively, and averaging over the small vertical scale $z$ to obtain \citep[see also][]{mC13}
\begineq
\dst \lp \psi_0' + J(\psi_0',\lp \psi_0') - \dz \pfw_0 =  \hz \cdot \nabla \times \lb \mBb_0 \cdot \nabla_\perp \langle \bbp_{1/2} \rangle \rb + \frac{1}{Re} \nabla_\perp^4 \psi_0' , \label{E:vort0}
\endeq
\begineq
\dst \pfw_0 + J(\psi_0',\pfw_0) +  \dz \psi_0' = \wG \langle \pth_1 \rangle + \hz \cdot \lb \mBb_0 \cdot \nabla_\perp \langle \bbp_{1/2} \rangle  \rb + \frac{1}{Re} \lp \pfw_0 ,\label{E:mom0}
\endeq
where $J(F,G) = \dsx F \dsy G - \dsx G \dsy F$, the angled brackets denote a spatial average over $z$, and the vertical vorticity is $\zeta_0 = \lp \psi_0'$.

Upon rescaling the velocity with the small-scale viscous diffusion time such that $U = \nu/L$, the closed set of reduced equations is given by
\begin{gather}
  \mub_{1/2} =  - \overline{\nabla} \times \mPsi_{1/2} \hz ,\label{E:mgeo2}\\
  \dz \mPsi_{1/2} =  \frac{\Rat}{Pr} \mth_0 ,\label{E:mhydro2} \\ 
  \dt \mth_0 + \mub_{1/2} \cdot \nabla_X \mth_0 + \dz \overline{\lb \pfw_0 \pth_1 \rb} =  \frac{1}{Pr} \dzt \mth_0 , \label{E:mheat} \\
  \partial_\tau \mBb^{\perp}_0 = \dz \lsq \hz \times \overline{\lb \ubp_0 \times \langle \bbp_{1/2} \rangle  \rb} \rsq + \frac{1}{\Pmt} \dzt \mBb^{\perp}_0, \label{E:minduc}\\
  \partial_\tau \mBz_{1/2} = \dx \lsq \hy \cdot \overline{\lb \ubp_0 \times \langle \bbp_{1/2} \rangle  \rb} \rsq - \dy \lsq \hx \cdot \overline{\lb \ubp_0 \times \langle \bbp_{1/2} \rangle  \rb} \rsq  + \frac{1}{\Pmt} \dzt \mBz_{1/2},\label{E:minducZ} \\
   \dx \mBx_0 + \dy \mBy_0 + \dz \mBz_{1/2} = 0, \label{E:mBSol}\\
   \dst \lp \psi_0' + J(\psi_0',\lp \psi_0') - \dz \pfw_0 =  \hz \cdot \nabla \times \lb \mBb_0 \cdot \nabla_\perp \langle \bbp_{1/2} \rangle \rb + \nabla_\perp^4 \psi_0', \label{E:fvort} \\
   \dst \pfw_0 + J(\psi_0',\pfw_0) +  \dz \psi_0' = \frac{\Rat}{Pr} \langle \pth_1 \rangle + \mBb_0 \cdot \nabla_\perp \langle \pbz_{1/2} \rangle + \lp \pfw_0, \label{E:fmom} \\
   \dst \langle \pth_1 \rangle + J(\psi_0',\langle \pth_1 \rangle) +  \pfw_0 \dz \mth_0 = \frac{1}{Pr} \lp \langle \pth_1 \rangle, \label{E:fheat} \\
   0 = \mBb^{\perp}_0 \cdot \nabla_\perp \ubp_0 + \frac{1}{\Pmt} \lp \langle \bbp_{1/2} \rangle ,\label{E:finduc}  \\
  \dsx \langle \pbx_{1/2} \rangle  + \dsy \langle \pby_{1/2} \rangle  = 0. \label{E:fBSol}
\end{gather}
The reduced Rayleigh number, consistent with the linear theory of rapidly rotating convection \citep{sC61}, is defined by $\Rat =\ep^4 Ra_H = E_H^{4/3} Ra_H$, where the Rayleigh number is given by
\be
Ra_H = \frac{g \alpha \Delta T H^3}{\nu \kappa}.
\ee
In addition, the reduced magnetic Prandtl number appearing in the above system is denoted by $\Pmt$.

The final reduced system is 10th order in the large-scale vertical coordinate $Z$.  Written in terms of the reduced variables, the ten boundary conditions become 
\be
\mth_0 = 1 \quad \textnormal{at} \quad Z = 0, \qquad \mth_0 = 0 \quad \textnormal{at} \quad Z = 1 , \label{E:bound1}
\ee
\be
\dz \mBb^{\perp}_0 = \mBz_{1/2} = \pfw_0 =  0 \quad \textnormal{at} \quad Z = 0,1. \label{E:bound2}
\ee

Although the details of obtaining numerical solutions to the reduced system are beyond the scope of the present work, we mention that several approaches have been developed to handle both multiple scales in space and time.  In particular, so-called heterogeneous multiscale methods (HMM) have been successfully applied to numerous applications \citep{wE07}.  For example, a modifed form of HMM has recently been developed for simulating a multiscale model of upper ocean Langmuir circulation \citep{gC09,zM14}. In addition, \cite{tH14} have developed a time-stepping algorithm for problems with temporal scale separation which may be of use for the present model.  

\subsubsection{The $Rm = O(1)$ Limit}

The asymptotic development of the $Rm = O(1)$ parallels that for the $Rm = O(\ep^{1/2})$ case given above.  For this reason we omit many of the  details and focus only on the differences between the two limiting cases.  We emphasize again that the $Rm=O(1)$ limit corresponds to $Pm=O(1)$ since $Re=O(1)$.  Consideration of relationships \eqref{E:Bt} and \eqref{E:Ms} with $Rm = O(1)$ shows that we must have
\be
A_\tau = \ep^{-1}, \quad M = O(1) . \label{E:Rm1D}
\ee
In addition, relationship \eqref{E:Bs1} shows that the mean and fluctuating magnetic fields are now of the same order, i.e.~$\mBb \sim \bbp$.  The main difference that occurs in the final reduced model for the $Rm = O(1)$ case is the form of the induction equations, which at leading order become
\begin{gather}
  \partial_\tau \mBb^{\perp}_0 = \dz \lsq \hz \times \overline{\lb \ubp_0 \times \langle \bbp_0 \rangle  \rb} \rsq , \label{E:minduc2}\\
  \partial_\tau \mBz_{1/2} = \dx \lsq \hy \cdot \overline{\lb \ubp_0 \times \langle \bbp_{0} \rangle  \rb} \rsq - \dy \lsq \hx \cdot \overline{\lb \ubp_0 \times \langle \bbp_{0} \rangle  \rb} \rsq  ,\label{E:minducZ2} \\
   \dst \langle \bbp_0 \rangle + \ubp_0 \cdot \nabla_\perp \langle \bbp_0 \rangle = \mBb^{\perp}_0 \cdot \nabla_\perp \ubp_0 + \frac{1}{Pm} \lp \langle \bbp_{0} \rangle .\label{E:finduc2}
\end{gather}
The fluctuating equations then take a mathematically equivalent form of equations \eqref{E:fvort}-\eqref{E:fheat} and \eqref{E:fBSol} by making the substitution $\langle \bbp_{1/2} \rangle \rightarrow  \langle \bbp_0 \rangle$.  

The boundary conditions for the $Rm=O(1)$ case are identical to those given by \eqref{E:bound1}-\eqref{E:bound2}.  Equations \eqref{E:minduc2}-\eqref{E:minducZ2} shows that due to the absence of $Z$-derivatives, no mean magnetic field boundary conditions can be satisified without the inclusion of magnetic boundary layers. Future work is necessary to examine the effects of these boundary layers.    


\section{Discussion}
\label{S:discuss}

\subsection{The $Rm=O(\ep^{1/2})$ limit}

Here we reiterate some of the key features of the low magnetic Prandtl number QGDM and interpret the various terms in the equation set \eqref{E:mgeo2}-\eqref{E:fBSol}.  Equations \eqref{E:mgeo2}-\eqref{E:mhydro2} are statements of geostrophic balance and hydrostatic balance on the large horizontal and vertical scales, respectively.  Importantly, these equations are diagnostic since they contain no information about the temporal evolution; this is a well-known characteristic and the prognostic dynamics are obtained from the mean heat equation \eqref{E:mheat}.  From equation \eqref{E:mgeo2} we see that viscous diffusion is negligible on the large scales.  The mean heat equation given by \eqref{E:mheat} shows that the mean velocity field is strong enough to allow advection of the mean temperature over the large horizontal scales $(X,Y)$ as shown by the second term on the lefthand side, whereas the third term represents the influence of convective feedback from the fluctuating scales back onto the large-scales.  Additionally, the presence of large-scale thermal diffusion over the vertical dimension is represented by the term on the righthand side of equation \eqref{E:mheat}.   

Both of the mean induction equations given by \eqref{E:minduc} and \eqref{E:minducZ} contain time dependence, the mean emf that represents the feedback from fluctuating velocity and magnetic field dynamics, and ohmic diffusion on the large vertical scale $Z$.  Thus, ohmic dissipation dominates viscous dissipation on the large scales since, at this order of approximation, viscous diffusion is not present in equations \eqref{E:mgeo2}-\eqref{E:mhydro2}.  This feature is consistent with the $Pm \ll 1$ limit considered here. Moreover, the QGDM is characterized by a horizontal magnetic field that is $O(Rm^{-1})$ stronger than the vertical magnetic field.    

Equations \eqref{E:fvort}-\eqref{E:fheat} are a magnetic version of the quasi-geostrophic convection equations originally developed by \cite{mS06} and \cite{kJ06}.  An important feature of equations \eqref{E:fvort}-\eqref{E:fmom} is that both advection and diffusion only occur over the small horizontal scales $(x,y)$ due to the anisotropic spatial structure of low Rossby number convection.  The fluctuating vorticity equation \eqref{E:fvort} contains time dependence, advection on the small-scales, vortex stretching is represented by the third term on the lefthand side, with the remaining terms on the righthand side being the Lorentz force and viscous diffusion, respectively. The fluctuating momentum equation is given by \eqref{E:fmom} and, like the fluctuating vorticity equation, contains time dependence and horizontal advection.  The vertical pressure gradient is given by the third term on the lefthand side of equation \eqref{E:fmom}, with the fluctuating geostrophic stream function $\psi_0'$ acting as pressure since the flow is geostrophically balanced; the remaining terms on the righthand side are the buoyancy force, the Lorentz force, and viscous diffusion. The fluctuating heat equation \eqref{E:fheat} is also characterized by time dependence and horizontal advection and diffusion, with the third term on the lefthand side representing advection of the mean heat by the fluctuating vertical velocity over the large-scale $Z$.  

The fluctuating induction equation \eqref{E:finduc} is identical to the equation given in S74, where he noted that the fluctuating magnetic field is induced by stretching the mean magnetic field with the small-scale strain $\nabla_\perp \ubp_0$ (consistent with our low $Rm$ approximation).  The absence of the time derivative shows that the fluctuating magnetic field adjusts instantaneously relative to the fluctuating convection, showing that magnetic diffusion is much more rapid than momentum diffusion on the small-scales; this effect is consistent with a small magnetic Prandtl number limit.  Alfv\'en waves are therefore damped out on the small horizontal scales.  As with the fluctuating vorticity, momentum, and heat equations, only horizontal diffusion is present in the fluctuating induction owing to spatial anisotropy. Taken with equations \eqref{E:fvort}-\eqref{E:fheat}, we see that both viscous dissipation and ohmic dissipation are important features for the small-scale dynamics.  The small-scale induction equation also shows that both the present model and the CSDM require the presence of a non-trivial mean magnetic field for the development of a dynamo; this shows that the resulting dynamo is therefore large-scale and typical of low $Rm$ dynamos \citep{hM78b}.

\subsection{The $Rm=O(1)$ limit}

The $Rm=O(1)$ case is associated with an order one magnetic Prandtl number.  For this reason, the $Pm=O(1)$ QGDM may be particularly important for relating to DNS studies where reducing $Pm$ to physically realistic values is impossible given the modest Reynolds numbers attainable with current computational resources.  Given that the energy and momentum equations are identical to the low $Pm$ QGDM discussed in the previous section, we focus here on the differences associated with the form of the induction equations \eqref{E:minduc2}-\eqref{E:finduc2}.    

The distinguished limit $A_\tau=\ep^{-1}$ shows that, in comparison to the $Rm=O(\ep^{1/2})$ case, the mean magnetic field now evolves on a faster timescale.  The $M=O(1)$ limit indicates that the magnetic and kinetic energies are now equipartitioned when $Pm=O(1)$ -- a result that appears to be consistent with DNS investigations \citep{sS04}.  

The absence of $Z$-derivatives in equations \eqref{E:minduc2}-\eqref{E:minducZ2} shows that ohmic dissipation is weak on the large-scales and limited to magnetic boundary layers adjacent to the horizontal boundaries.  The small-scale induction equation \eqref{E:finduc2} contains time dependence, advection by the fluctuating velocity field, along with stretching and diffusion.  The presence of $\dst \langle \bbp_0 \rangle$ results in a fluctuating magnetic field that no longer adjusts instantaneously to the fluctuating velocity field as it does when $Pm \ll 1$.  For this case, Alfv\'en waves are present on the small fluctuating scales.


\subsection{Relationship with the Childress-Soward dynamo model (CSDM)}

In Table \ref{T:lims} we show the different distinguished limits taken between the present work and those taken by S74 and FC82.  Although we have considered two limits of $Rm$ that result in different limits of $M$, $Pm$, and $A_\tau$, the present discussion will be focused on the $Rm=O(\ep^{1/2})$ (low $Pm$) QGDM.  Both the weak and intermediate field forms of the CSDM can be derived directly from equations \eqref{E:meanmom}-\eqref{E:flucsolen} by employing the distinguished limits listed under the columns labeled ``Soward" and ``Fautrelle \& Childress", respectively, and taking $\dst = \dt = \dx = \dy \equiv 0$.  The consequence of taking $Re = Pe = O(\ep^{1/2})$ is that the resulting models are weakly nonlinear since (horizontal) viscous diffusion enters at higher order than the advective nonlinearities in the fluctuating momentum equation.  While in the present work we have considered both low and order one $Pm$, the CSDM employs $Pm = O(1)$.  Additionally, the magnetic field scaling $M^{S}=\ep^{3/2}$ employed by S74 and $M^{FC}=\ep^{1/2}$ employed by FC82 contrasts sharply with our $M=\ep^{-1/2}$ limit; the result is that magnetic energy is significantly larger than the kinetic energy in the present work.

\begin{table}
  \begin{center}
    \begin{tabular}{cccc}
      Parameter   \vline \quad  &    Present Work  \vline \quad&    Soward  \vline \quad & Fautrelle \& Childress  \\
      \hline
      $Eu$     &    $\ep^{-2}$     &    $\ep^{-2}$   &    $\ep^{-2}$  \\  
      \vspace{0.1cm}    
      $Re$      &    $O(1)$     &    $\ep^{1/2}$  &    $\ep^{1/2}$  \\  
       \vspace{0.1cm}  
      $M$     &    $\ep^{-1/2}, O(1)$     &    $\ep^{3/2}$   &    $\ep^{1/2}$ \\
       \vspace{0.1cm} 
      $\Gamma$     &    $\wG \ep^{-1}$     &    $\wG \ep^{-1}$   &    $\wG \ep^{-1}$ \\
       \vspace{0.1cm} 
       $Pe$      &    $O(1)$     &    $\ep^{1/2}$   &    $\ep^{1/2}$ \\ 
        \vspace{0.1cm} 
      $Rm$     &    $\ep^{1/2}, O(1)$     &    $\ep^{1/2}$    &    $\ep^{1/2}$ \\
       \vspace{0.1cm} 
      $Pr$     &    $O(1)$     &    $O(1)$    &    $O(1)$  \\
       \vspace{0.1cm} 
      $Pm$     &    $\ep^{1/2}, O(1)$     &    $O(1)$   &    $O(1)$  \\  
       \vspace{0.1cm}     
      $A_X$     &    $\ep^{-3/2}$     &    $-$   &    $-$  \\
       \vspace{0.1cm} 
      $A_Z$     &    $\ep^{-1}$     &    $\ep^{-1}$  &    $\ep^{-1}$  \\
       \vspace{0.1cm} 
      $A_\tau$     &    $\ep^{-3/2}$     &    $\ep^{-3/2}$   &    $\ep^{-3/2}$ \\
       \vspace{0.1cm} 
      $A_T$     &    $\ep^{-2}$     &    $-$   &    $-$ \\
    \end{tabular}
    \caption{Comparison between the different distinguished limits taken in the present work and those of \cite{aS74} and \cite{yF82}, relative to equations \eqref{E:meanmom}-\eqref{E:flucsolen}.  We note that both Soward and Fautrelle \& Childress considered weakly nonlinear motions and only a single (slow) timescale and a single large-scale coordinate in the governing equations such that $\dst = \dt = \dx = \dy \equiv 0$.  In all models the small parameter is the Rossby number, $Ro = \ep$, and the reduced buoyancy number $\wG = O(1)$.}
    \label{T:lims}
  \end{center}
\end{table}

CS72 described three different classes of dynamos, distinguished by the relative strength of the magnetic field. In their work, the classification was based on the magnitude of the large-scale Hartmann number defined as
\be
Ha_H = \frac{\mathcal{B} H}{\lb \mu \rho \nu \eta \rb^{1/2}}. 
\ee
In the CS72 terminology, the weak field, intermediate field, and strong field dynamos are characterized by Hartmann numbers of magnitude $Ha^{w}_H = O(1)$, $Ha^{i}_H = O(\ep^{-1/2})$ and $Ha^{s}_H \gg \ep^{-1/2}$, respectively.  FC82 more precisely identified the strong field regime by the scaling $Ha^{s}_H = \ep^{-3/2}$ since this results in a Lorentz force that is comparable in strength to the Coriolis force.  By employing our distinguished limits, the large-scale Hartmann number in the present work is given by 
\be
Ha^{*}_H = \frac{\lb M Rm \rb^{1/2}}{\ep} = \frac{1}{\ep}.
\ee
Thus, by the classifications of CS72 and FC82 the present model considers magnetic fields with strengths that lie between the intermediate and strong field limits.  As the fluctuating momentum equation \eqref{E:momfluc} shows, both the Lorentz force and horizontal viscous diffusion enter the prognostic equation at the same order in our model; the result is that the small-scale Hartmann number is unity, $Ha^* = 1$.  

A strong field dynamo is synonomous with the magnetostrophic balance, whereby the Coriolis, pressure gradient and Lorentz forces are of comparable magnitude.  Our present model is geostrophically balanced on both the large and small scales, though as equation \eqref{E:mag} shows a magnetostrophic balance does appear at higher, but subdominant, order in the mean momentum equation.  A commonly employed dimensionless measure of the strong field dynamo is the Elsasser number, defined as the ratio of the Lorentz force to the Coriolos force and often expressed as
\be
\Lambda = \frac{\mathcal{B}^2 }{2 \Omega \rho \mu \eta} .\label{E:El}
\ee
In terms of the nondimensional parameters and distinguished limits employed in the present study (e.g. Table \ref{T:lims}) we have
\be
\Lambda^* = M Pm Ro Re = \ep.
\ee
Thus, as expected, the Elsasser number is small in our geostrophically balanced model.  The above scaling is also consistent with previous quasi-geostrophic magnetoconvection studies \citep[c.f.][]{cJ03b,nG07b}.  In S74 and FC82 the Elsasser number is significantly smaller with $\Lambda^S = \ep^3$ and $\Lambda^{FC} = \ep^2$, respectively.  Although DNS studies are often characterized by order one Elsasser number, we note that these models are all limited to $Pm \sim O(1)$ and moderate values of the Ekman number. We speculate that if these models could increase their magnetic Reynolds number solely by increasing their Reynolds number and decreasing the Ekman number, rather than increasing $Pm$, they would begin to reach significantly smaller Elsasser numbers; this trend may be apparent in the plane layer DNS investigation of \cite{sS04} and the spherical DNS study of \cite{uC06}.  Also, we note that the value of the Elsasser number is also dependent upon how one scales the magnetic field. The form given by \eqref{E:El} is dependent upon the magnetic diffusivity, and independent of the velocity and length scales that are important for assessing dynamical balances; this is because it is typically assumed that the magnitude of the current density scales as $\mathcal{J} \sim U \mathcal{B}/\mu \eta$ \citep[e.g.][]{pD01}.  To be consistent, the Elsasser number should be defined directly from the governing equations as 
\be
\widetilde{\Lambda} = \frac{\frac{1}{\rho \mu} \Bb \cdot \nabla \Bb}{2 \Omega\hz \times \ub} = \frac{\mathcal{B}^2 }{2 \Omega \rho \mu l \mathcal{U}}. \label{E:El2}
\ee
In the present work we have
\be
\widetilde{\Lambda}^* = M Ro = \ep^{1/2} . \label{E:El3}
\ee
Similarly, for the $Rm=O(1)$ QGDM, this gives $\widetilde{\Lambda}^*=\ep$.  This shows that although the Elsasser number is still small, it is independent of the magnetic diffusivity. Given that the model developed in the present work is characterized by small Elsasser number, yet the magnetic energy is asymptotically larger than the kinetic energy for the low $Pm$ QGDM, we can generally say that the partitioning of magnetic and kinetic energy is not necessarily indicative of dominant balances present within the governing equations. 

Finally, we note that it is possible to extend the CSDM to include fully nonlinear motions, a strong magnetic field, and small $Pm$ by neglecting the large-scale horizontal modulations (i.e.~$\dx = \dy = 0$) appearing in equations \eqref{E:mhydro2}-\eqref{E:fBSol}; the result is given by
\begin{gather}
  \dz \overline{\Psi}_{1/2} =  \frac{\Rat}{Pr} \mth_0 ,\\ 
\dt \mth_0 + \dz \overline{\lb \pfw_0 \pth_1 \rb} =  \frac{1}{Pr} \dzt \mth_0 , \\
 \partial_\tau \mBb^{\perp}_0 = \dz \lsq \hz \times \overline{\lb \ubp_0 \times \langle \bbp_{1/2} \rangle  \rb} \rsq + \frac{1}{\Pmt} \dzt \mBb^{\perp}_0, \\
\dst \lp \psi_0' + J(\psi_0',\lp \psi_0') - \dz \pfw_0 =  \hz \cdot \nabla \times \lb \mBb_0 \cdot \nabla_\perp \langle \bbp_{1/2} \rangle \rb + \nabla_\perp^4 \psi_0', \\
\dst \pfw_0 + J(\psi_0',\pfw_0) +  \dz \psi_0' = \frac{\Rat}{Pr} \langle \pth_1 \rangle + \mBb_0 \cdot \nabla_\perp \langle \pbz_{1/2} \rangle + \lp \pfw_0, \\
 \dst \langle \pth_1 \rangle + J(\psi_0', \langle \pth_1 \rangle) +  \pfw_0 \dz \mth_0 = \frac{1}{Pr} \lp \langle \pth_1 \rangle, \\
  0 = \mBb^{\perp}_0 \cdot \nabla_\perp \ubp_0 + \frac{1}{\Pmt} \lp \langle \bbp_{1/2} \rangle ,\\
\dsx \langle \pbx_{1/2} \rangle  + \dsy \langle \pby_{1/2} \rangle  = 0,
\end{gather}
and we note that equaton \eqref{E:mBSol} is trivially satisfied in this case.

\subsection{Applications to natural dynamos}
\label{S:apps}

It is informative to compare the distinguished limits that were taken in the present work with what is known about the geodynamo and other natural dynamos;  this exercise is useful for establishing the strengths and weaknesses of the present model.  In the present work three timescales were necessary to allow time variations of the fluctuating convection, the mean magnetic field, and the mean temperature.  The relative ordering of these timescales is given by 
\begineq
t \ll A_\tau \tau \ll A_T T , \label{E:time}
\endeq
or
\begineq
t \ll \ep^{-3/2} \tau \ll \ep^{-2} T ,
\endeq
upon noting that $\tau = A_\tau^{-1} t$ and $T = A_T^{-1} t$.  For instance, an order one change in the slow timescale $\tau$ is asymptotically larger than an order one change in the fast timescale $t$.  This ordering states that the large-scale magnetic evolution timescale lies midway between the small-scale convective timescale and the large-scale thermal evolution timescale.  For the geodynamo, observations of the geomagnetic field show that $t \sim O(10^2)$ years, $\tau \sim O(10^4)$ years and $T \sim O(10^9)$ years, suggesting that the above ordering is realistic.  

  Utilizing the velocity based on the small-scale viscous diffusion timescale, the (small-scale) Rossby number can be related to the large-scale Ekman number as
\be
\ep = Ro = E_H^{1/3} .
\ee
Studies suggest that the viscosity of the Earth's core is similar to that of water at standard temperature and pressure \cite[e.g.][]{mP13} such that $E_H \sim O(10^{-15})$.  Estimates for the small- and large-scale Rossby numbers, and the large-scale magnetic and hydrodynamic Reynolds numbers are given by, respectively,
\begin{gather}
\ep  =  10^{-5},  \label{E:Ross} \\
Ro_H  = \ep^2 = 10^{-10}, \\
Rm_H  =  Rm E_H^{-1/3} = \ep^{-1/2} \sim O(10^2), \\
Re_H   =  Re E_H^{-1/3} = 10^5 Re.
\end{gather}
Although the small-scale Rossby number in the Earth's outer core is not known, the above estimate for the large-scale Rossby number is not too different from estimates based on observations \citep{cF11}.  The large-scale magnetic Reynolds number possesses the correct order of magnitude estimate for the core.  For the large-scale Reynolds number $Re_H$ we require knowledge of the currently unknown small-scale Reynolds number $Re$.  Numerical simulations of the non-magnetic quasi-geostrophic convection equations can currently attain values of $Re \lesssim O(10^3)$ \citep[e.g.][]{kJ12}, leading to large-scale Reynolds numbers of $Re_H \sim O(10^8)$; this suggests the QGDM is in the appropriate dynamical regime necessary for understanding the geodynamo. 

The current model assumes an order one small-scale Reynolds number; coupled with the relationship $Rm = Re Pm$, this implies that $Pm = O(\ep^{1/2})$.  The magnetic Prandtl number in the Earth's core is thought to be $Pm \sim O(10^{-6})$, so it would seem that the distinguished limit taken for $Pm$ is not quite in line with what typical values are in the Earth's core.  However, the theory developed here suggests that $Pm=O(\ep^{1/2})$ is sufficiently small to be in the asymptotic limit of small magnetic Prandtl number; we can then reach smaller magnetic Prandtl numbers by reducing $\Pmt$.  Furthermore, the distinguished limit of $M = O(\ep^{-1/2})$ shows that magnetic energy dominates when $Pm \ll 1$; this result is thought to be generally consistent with what is known about the geodynamo.  

As previously mentioned, we have chosen to non-dimensionalize the equations based on those dimensional quantities that possess the maximum amplitude; this implies that some non-dimensional parameters will contain a `mix' of large- and small-scale dimensional quantities.  Because of the importance of the small-scale motions, the dimensional length and velocity scales are then the small scales $l$ and $\mathcal{U}$, respectively.  Pressure, temperature, and magnetic field were then based on the large-scale mean quantities.  Any dimensionless parameter can be rescaled to be only a function of the small-scale or large-scale dimensional quantities by noting the asymptotic relations that were employed in the present work.  For instance, the magnetic to kinetic energy ratio $M$ is based on the large-scale magnetic field and the small-scale velocity field.  When expressed solely in terms of the large-scale variables we have $\overline{M} = \ep^{-1} M \approx 10^7$ where we've used the $\epsilon$ estimate given by \eqref{E:Ross}.

Assuming a large-scale magnetic field strength of $O(10^{-3})$ Tesla for the core \citep[e.g.][]{nG10}, the traditionally defined Elsasser number \eqref{E:El} is $\Lambda=O(1)$.  The model presented here is not capable of matching this value and represents one of the most significant discrepancies between the present work and what is presently known about the geodynamo.  With the exception of Jupiter, however, estimates suggest that the Elsasser number for other planetary dynamos is significantly less than unity \citep{gS11}.  As mentioned previously, the traditionally defined Elsasser number may not be an accurate indicator of the dominant force balance in the core as numerical simulations show that dynamos characterized by $\Lambda=O(1)$ remain geostrophically balanced \citep{kS12}.  With this result in mind we emphasize that some care must be taken with regard to estimating the dominant balance in DNS. Taking the curl of the momentum equation appears to be a common approach for determining the dominant force balance in simulations \citep{uC06,eK13b}.  Such a procedure can often lead to an erroneous conclusion, since the action of curling the momentum equation will automatically filter a dominant geostrophic balance. This is indeed the case for the present model where leading order geostrophic balances identified in equations \eqref{E:mgeo} and \eqref{E:flucgeo} are removed by the curling operation.  For instance, equation \eqref{E:mcont} shows that taking the curl of the mean momentum equation \eqref{E:mgeo} yields the trivial result up to $O(\ep^{1/2})$.  The first non-trivial result arises at $O(\ep)$ by curling the mean momentum equation \eqref{E:mag} and will yield a vorticity equation involving a (vorticity) balance between the curl of the Coriolis and Lorentz force terms; this should not be interpreted as a force balance between the Coriolis and Lorentz forces.   The salient point is that dominant balances can only be determined directly from the momentum equation, and not from the vorticity equation.  Rather, analysis of the terms present in the vorticity equation for a geostrophically balanced flow will tell you which forces act as \textit{small} perturbations to the main geostrophic balance.

The alternative definition of the Elsasser number given by equation \eqref{E:El2}, is $O(10^{-2})$ for the geodynamo, suggesting that the Lorentz force may not enter the leading order force balance for the large-scale dynamics of the core \citep{hN15}.  Moreover, previous work calculating the angular momentum of the Earth's outer core must invoke geostrophy to obtain the large-scale velocity field within the core, and shows excellent agreement with values obtained independently from observations of the Earth's rotation rate \citep{dJ88,aJ93,nG10,nS11,nG12}.  These results suggest that a large-scale thermal wind model, similar to the one developed here, may be relevant for understanding the physics of the geodynamo.  Of course, the large-scale thermal wind model is characterized by $A_X \gg A_Z$ and is therefore unlike that of a deep spherical shell where the aspect ratio is order one.  We were unable to obtain a closed large-scale model with $A_X = A_Z$ that is coupled to the small-scale dynamics; future work is necessary to determine if this difficulty can be circumvented in a spherical geometry.

The small-scale dynamics of the geodynamo are completely unconstrained due to lack of observations.  The present work has utilized an expansion based on the smallness of the small-scale Rossby number.  As discussed by \cite{hN15}, the precise value of the small-scale Rossby number depends upon the detailed properties of the turbulence that is controlled by the relative importance of inertia, buoyancy and the small-scale magnetic field.  \cite{hN15} argue that although the Elsasser number, as given by equation \eqref{E:El2}, is $O(0.01)$ for the largest scales of the Earth's core, it may increase to $O(1-10)$ on the scale of about $10$km such that the Lorentz force becomes more significant as the length scale is reduced.  Indeed, as the QGDM shows, if the Lorentz force is subdominant on the largest scales then it must be present in the small-scale dynamics for the dynamo to reach a saturated, non-kinematic state. Future numerical simulations of the QGDM should therefore help to understand the small-scale dynamics of the geodynamo and other planetary magnetic fields \citep[e.g.][]{sS10}.  

We emphasize that exact agreement between the above estimates for the non-dimensional parameters and observations of natural dynamos is not required for the asymptotic theory presented here to accurately model convection-driven dynamos.  As long as $\ep$ is small, the QGDM represents an accurate reduced form of the full set of governing equations and can faithfully model geostrophically balanced dynamos in the plane layer geometry.  For atmospheric and oceanic flows, for instance, it is known that QG theory remains accurate for Rossby numbers just above $\ep=O(0.1)$ \citep[e.g.][]{jP87}.  

\section{Conclusion}
\label{S:conclude}

In the present work we have utilized a standard multiple scales asymptotic approach to develop a new multiscale dynamo model that is valid in the limit of small Rossby number.  The small-scale model is characterized by a magnetically-modified version of the quasi-geostrophic convection equations originally developed by \cite{kJ98a} and later studied in detail by \cite{mS06}.  The large-scale model is characterized by a thermal wind balance in the mean momentum equations; coupled with the mean heat equation, the large-scale model is equivalent to the well-known planetary geostrophy equations commonly employed for investigating the dynamics of the Earth's atmosphere and ocean.  We have discussed both low and order one magnetic Prandtl number models, showing that these two cases possess fundamentally different properties.  For the low $Pm$ case the magnetic energy dominates the kinetic energy, and ohmic dissipation is asymptotically dominant over viscous dissipation on the large-scales. When $Pm = O(1)$ it was demonstrated that the magnetic and kinetic energies become equipartitioned with weak large-scale ohmic dissipation.  In both cases the dynamics are characterized by small Elsasser number since the motions are geostrophically balanced.  The new model can be considered a fully nonlinear, generalized version of the dynamo model originally developed by \cite{sC72}.

Numerical simulations of asymptotically reduced equation sets have proven useful for accessing dynamical regimes in rotating plane layer convection that are computationally demanding or impossible to reach with the use of DNS \citep[e.g.][]{mS06,kJ12b,kJ12,aR14,sS14}.  These investigations have shown the dependence of the flow regime on both the Rayleigh and Prandtl numbers, new phenomena such as an asymptotic heat transfer scaling regime \citep{kJ12b}, and large-scale vortex formation via an inverse cascade \citep{kJ12}.  Future simulations of the new model will help to shed light on the interaction of these phenomena with a magnetic field.  For instance, a recent DNS investigation by \cite{cG15} has shown that large-scale vortices can play a significant role in generating large-scale magnetic fields, particularly as the magnetic Prandtl number is reduced below one. 

Various extensions of the present work can also be carried out.  While we have assumed, for simplicity, that the fluid is Boussinesq and that the gravity vector and rotation vector are aligned, it is a straightforward procedure to relax both of these constraints to include compressibility and varying angle between the gravity and rotation vectors \citep{kJ06,kM13}.  Although the anelastic approximation appears to agree well with the compressible equations for the case of order one Prandtl numbers \citep{mC14}, it has recently been shown that it yields spurious results for low Prandtl number quasi-geostrophic convection \citep{mC15}.  However, \cite{mC15} outlined an approach for extending the Boussinesq quasi-geostrophic convection equations of \cite{mS06} to the case of a fully compressible gas. 

The present model has focused on the plane layer geometry for the sake of mathematical and physical simplicity.  To further the applicability of the present work with that of natural systems, it will be useful to extend the present methodology to the rotating cylindrical annulus \citep{fB86} and to spherical geometries.  The three-dimensional cylindrical annulus model recently developed by \cite{mC13} is particularly interesting since it possesses order one axial velocties, such that the Ekman pumping dynamo effect investigated by \cite{fB75} is no longer a prerequisite for dynamo action in this geometry.

\section*{Acknowledgements}
The authors thank Stephen Childress, three anonymous referees, and Meredith Plumley for helpful comments that improved the manuscript greatly. This work was supported by the National Science Foundation under grants EAR \#1320991 (MAC, KJ and JMA) and EAR CSEDI \#1067944 (KJ and JMA).

\bibliographystyle{jfm}

\bibliography{Dynamobib}

\end{document}